\documentclass[journal]{IEEEtran}
\usepackage{graphicx}
\usepackage{epstopdf}
\usepackage{cite}
\usepackage{algorithm}
\usepackage{algorithmic}
\usepackage{amsmath}
\usepackage{bm,comment,color,amssymb}
\usepackage{stfloats}
\usepackage[colorlinks]{hyperref}
\begin{document}
	
	\title{Exploring the Impact of RIS on Cooperative NOMA URLLC Systems: A Theoretical Perspective}
	
	\author{Jianchao Zheng, Tuo Wu, Junteng Yao,\\ Chau Yuen, \emph{Fellow, IEEE}, Zhiguo Ding, \emph{Fellow, IEEE}, \\and   Fumiyuki Adachi, \emph{Life Fellow}, \emph{IEEE}
		
	 \thanks{(\textit{Corresponding author: Tuo Wu.})}
	 \thanks{J. Zheng is with the School of Computer Science and Engineering, Huizhou University, Huizhou 516000, China (E-mail: $\rm zhengjch@hzu.edu.cn$). T. Wu and C. Yuen are with the School of Electrical and Electronic Engineering, Nanyang Technological University, 639798, Singapore (E-mail: $\rm \{tuo.wu, chau.yuen\}@ntu.edu.sg$). J. Yao is with the Faculty of Electrical Engineering and Computer Science, Ningbo University, Ningbo 315211, China (E-mail:  $\rm yaojunteng@nbu.edu.cn$). Zhiguo Ding is with Department of Electrical Engineering and Computer Science, Khalifa University, Abu Dhabi 127788, UAE, and University of Manchester, Manchester, M1 9BB, UK. (E-mail:  $\rm zhiguo.ding@ieee.org$). F. Adachi is with the International Research Institute of Disaster Science (IRIDeS), Tohoku University, Sendai, Japan (E-mail: $\rm adachi@ecei.tohoku.ac.jp$).}   
		
	}
	
	\markboth{}
	{Zheng \MakeLowercase{\textit{et al.}}:  Exploring the Impact of RIS on Cooperative NOMA URLLC Systems: A Theoretical Perspective}
	
	\maketitle

	\begin{abstract}
In this paper, we conduct a theoretical analysis of how to integrate reconfigurable intelligent surfaces (RIS) with cooperative non-orthogonal multiple access (NOMA), considering URLLC. We consider a downlink two-user cooperative NOMA system employing short-packet communications, where the two users are denoted by the central user (CU) and the cell-edge user (CEU), respectively, and an RIS is deployed to enhance signal quality. Specifically, compared to CEU, CU lies nearer from BS and enjoys the higher channel gains. Closed-form expressions for the CU's average block error rate (BLER) are derived.  Furthermore, we evaluate the CEU's BLER performance utilizing selective combining (SC) and derive a tight lower bound under maximum ratio combining (MRC). Simulation results are provided to  our analyses and demonstrate that the RIS-assisted system significantly outperforms its counterpart without RIS in terms of BLER.  {Notably, MRC achieves a squared multiple of the diversity gain of the SC}, leading to more reliable performance, especially for the CEU. Furthermore, by dividing the RIS into two zones, each dedicated to a specific user, the average BLER can be further reduced, particularly for the CEU.
	\end{abstract}
	\IEEEpeerreviewmaketitle

	\section{Introduction}
	\subsection{Background}
	The proliferation of wireless technologies heralds a new era in wireless communications. Reconfigurable intelligent surfaces (RISs) \cite{WuQ21}, a pivotal technology for 6G, have garnered immense attention. They provide the new flexibility to configure the wireless propagation environment in order to a transmission medium with more favorable characteristics \cite{PanC22, ShiE24, ZhiK21, ZhiK22, RenH22, TWu2}.  Leveraging the capabilities of RISs, the industry stands on the brink of a revolution where link quality and coverage in wireless networks can be enhanced to unprecedented levels, unlocking a plethora of applications and novel research trajectories, as highlighted in \cite{PanC21}. Parallel to the rise of RISs, non-orthogonal multiple access (NOMA) is emerging with prominence, etching its mark as an instrumental technology for next-generation wireless communication. NOMA not only offers a seamless blend of scalability and adaptability but also ushers in tangible enhancements in energy and spectral efficiency, factors that are cardinal to the evolution of communication infrastructures \cite{DingZ14,IslamS16,JZheng24,DingZ16}.
	\subsection{Related Works}
	The fusion of RIS and NOMA technologies into a unified network framework has unveiled a compelling synergy of their inherent advantages. This amalgamation has piqued the curiosity of many researchers, leading to exploratory studies like those showcased in \cite{ZhengB20,PawarA23,ChengY21,ChengY212,ReddyP22,AsieduD23,BariahL21,LiS22}. A seminal work by Zheng \emph{et al.} \cite{ZhengB20} delves into a rigorous theoretical analysis comparing the efficacies of NOMA and its counterpart, orthogonal multiple access (OMA), specifically in scenarios bolstered by RIS in downlink communication. This examination underscores the challenges of formulating power minimization problems, especially within the constraints of discrete unit-modulus reflection for each RIS element. Building on this concept, later studies by authors in \cite{ChengY21} and \cite{ChengY212} employed the outage performance metrics of NOMA systems assisted by multiple RISs with discrete phase shifts. They further investigated both downlink and uplink dynamics of RIS-enhanced NOMA and OMA under the influence of Nakagami-$m$ fading channels. Recognizing the potential of large intelligent surfaces (LIS) to augment communication paradigms, Barish \emph{et al.} tabled a robust mathematical framework in \cite{BariahL21} to dissect the error rate metrics within an LIS-empowered NOMA infrastructure. Complementing this research arc, Li \emph{et al.} \cite{LiS22} delineated the outage probabilities inherent to downlink RIS-boosted backscatter communication integrated with NOMA.
	
	In light of the robust improvements by integrating RIS and NOMA technologies, another significant stride in the wireless communication arena has been the introduction of short-packet communications. Pioneered by Polyanskiy et al., this paradigm was introduced as a response to the urgent need for reduced communication latencys \cite{Polyanskiy10}. Block error rate (BLER) subsequently emerged as a pivotal performance metric, precisely capturing the performance of these short-packet transmissions \cite{Polyanskiy10,Makki14,YuY18}.  Envisioning a system that prossess both the efficiency of low communication latency and the quality of communication link, researchers delved into the amalgamation of RIS technologies with short-packet communications  \cite{SuiL23,DeshpandeR23,XieH21,YuanL23,ZhangB22}. For instance,  Sui \emph{et al.} \cite{SuiL23} formulated a lower bound, shedding light on the decoding error probabilities concerning the optimal codes, given a certain signal-to-noise ratio and a code rate, especially in the domain of RIS  communication systems operating over Rician fading channels within short blocklength regimes. Furthermore, Yuan \emph{et al.} \cite{YuanL23} presented an in-depth analysis of an RIS-enhanced short-packet communication architecture, specifically benchmarked against Nakagami-$m$ fading channels. Additionally, in \cite{ZhangB22}, the approximate closed-form expression of the average packet error probability  was developed and the channel uses in the wireless energy transfer and wireless information transfer phases were optimized to maximize the effective system throughput.
	
	In the quest to realize both minimal communication latency and enhanced spectral efficiency, the integration of short-packet communications within NOMA frameworks has been the subject of significant investigation \cite{YuY18,SunX18,LaiX19,XieX21,YaoJ22}. Specifically, Yu \emph{et al.} \cite{YuY18} delved into the performance metrics of a two-user NOMA system tailored for short-packet transmissions. Extending this exploration, Sun \emph{et al.} \cite{SunX18} focused on optimizing power allocation strategies, aiming to maximize the effective throughput in the NOMA-based short-packet architecture. Lai \emph{et al.} \cite{LaiX19} took this a step further by examining cooperative approaches within the NOMA short-packet realm, subsequently offering closed-form expressions for the average BLER. Finally, Yao \emph{et al.} \cite{YaoJ22} examined the optimization problem of decoding error probability and power allocation factors, focusing on maximizing the effective throughput at the central user (CU) while ensuring the cell-edge user (CEU) meets the minimum required effective throughput constraint. 
	\subsection{Motivations and Contributions}
Recent research has increasingly focused on integrating RIS, NOMA, and ultra-reliable low-latency communication (URLLC) to enhance wireless communication systems \cite{YuanL23,VuT22,XuJ23,FengL2020,WangJ22,YangJ22}. Vu et al. \cite{VuT22} derived closed-form expressions for the average BLER in RIS-aided short-packet NOMA systems, considering both random and optimal phase shifts, as well as scenarios with perfect and imperfect successive interference cancellation (SIC). Similarly, Xu et al. \cite{XuJ23} analyzed the performance of simultaneous transmissions in RIS-assisted NOMA URLLC communications, accounting for continuous and discrete phase shifts. However, traditional NOMA systems suffer from multi-user interference, which can degrade performance even in RIS-enhanced environments. This interference impacts RIS-NOMA short-packet transmissions and can lead to performance instability due to channel variations. To tackle these challenges, cooperative NOMA has been introduced, wherein stronger users serve as relays to support weaker users, reducing interference and improving stability \cite{DingZ15, LiY18, LaiX19}.

Motivated by these considerations, this paper explores the impact of RIS on cooperative NOMA URLLC systems from a theoretical perspective. Specifically, we investigate the average BLER in RIS-assisted cooperative NOMA URLLC communication systems to understand how RIS can enhance system performance. To optimize communication for both the CU and the CEU, we draw inspiration from \cite{BariahL21,LiS22} and propose to divide the RIS into two distinct zones, each dedicated to serving one user. This configuration aims to fully exploit the capabilities of RIS in mitigating inter-user interference and improving the reliability and latency of cooperative NOMA URLLC systems.
 By providing a theoretical analysis of the RIS's impact on cooperative NOMA URLLC systems, this paper yields a better understanding of how using RIS can be leveraged to overcome challenges in next-generation wireless communications. Our findings offer valuable insights for the design and optimization of advanced wireless networks that require ultra-reliability and low latency.
	
The key contributions of this paper are summarized as follows:
	\begin{itemize}
		 \item This paper considers a downlink cooperative NOMA URLLC system enhanced by a RIS, comprising a BS, two users, namely CU and CEU.  {Compared to CEU, CU lies nearer from BS and enjoys the higher channel gains.} For this setup,  the closed-form expressions for the average BLER of the CU using the probability density functions (PDFs) of the channels are derived.
		
		\item The average BLER for the CEU is derived with selective combining (SC) employed. Furthermore, a tight lower bound on the CEU's average BLER is obtained considering maximum ratio combining (MRC) is used, providing insights into the performance differences between SC and MRC in RIS-assisted cooperative NOMA systems.
		
		\item The diversity order of the proposed system is analyzed in the high signal-to-noise ratio (SNR) regime. By calculating the slopes of the BLER, it is found that {MRC achieves a squared multiple of the diversity gain of the SC}. This indicates that MRC provides more reliable performance, especially benefiting the CEU which suffers from challenging reception conditions.

		\item Through extensive simulations, we validate the accuracy of our theoretical analyses. The results demonstrate that the RIS-aided cooperative NOMA system significantly outperforms its counterpart without RIS in terms of BLER performance. The performance gains are especially pronounced for the CEU, particularly for the case with MRC and at high SNR.
		
	\end{itemize} 
	
The rest of this paper is organized as follows. Section II describes the RIS-assisted downlink cooperative NOMA communication system. In Section III, the theoretical analysis of the average BLER for the CU and CEU is provided. Section IV presents the diversity order in the high SNR regime. Simulation results are discussed in Section V. Finally, conclusions are drawn in Section VI.
	
	\section{System Model} 
\begin{figure}
	\centering
	\includegraphics[width=3.2in]{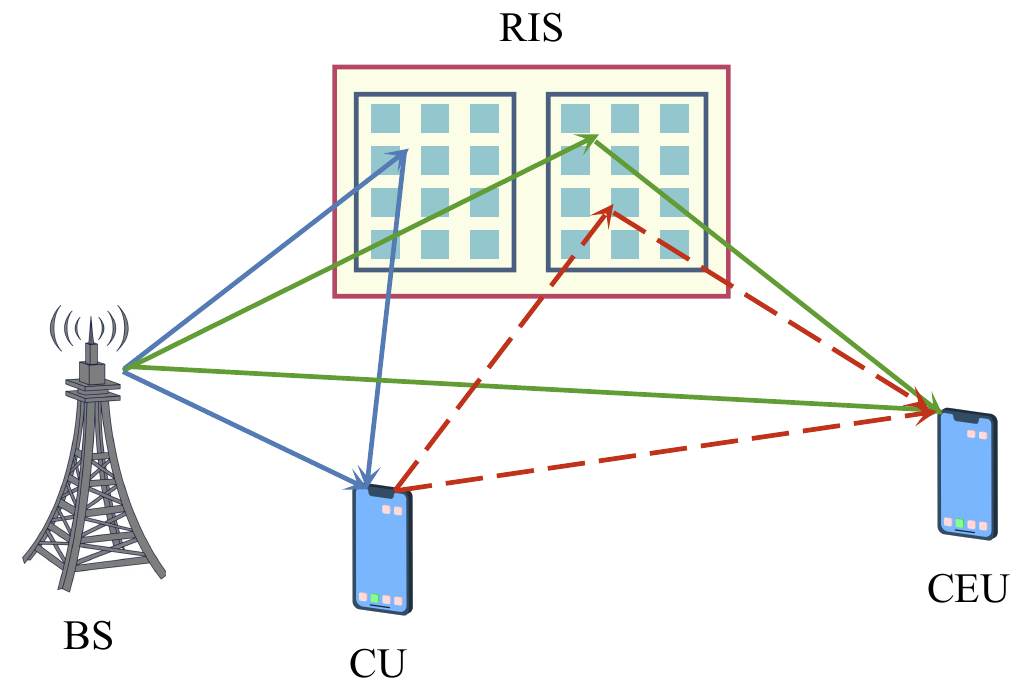}
	\caption{  RIS-aided cooperative NOMA URLLC systems.  {At the first phase, the BS transmits signals to two users based on NOMA scheme, which is indicated by blue solid lines and green solid lines. At the second phase, CU cooperatively relays the signals intended to CEU, which is indicated by red dashed lines.}} \label{sm}
\end{figure}

	As depicted in Fig. \ref{sm}, an RIS-assisted downlink cooperative NOMA system is studied, {where RIS has a perfect knowledge of the direction to BS, CU, and CEU.} We consider all the components, BS and the two users, are equipped with a single antenna. {Compared to CEU, we assume the the location of the CU is nearer from BS and enjoys the higher channel gains.} The BS aims to transmit $N_c$ and $N_e$ bits with a block length of $m$  to the CU and CEU, respectively, via an RIS. And we assume that there is no block between BS and the two users.
	
	To enhance communication performance, the RIS is segmented into two distinct zones, each employed with $R$ reflective elements \cite{BariahL21,LiS22}, as shown in Fig. 1. Furthermore, each zone is configured with unique phase shift capabilities to accurately direct the reflected signals to specific users. The first zone is tasked with reflecting signals from the BS to the CU, while the second zone directs signals to the CEU.
	
	The system utilizes a two-phase downlink NOMA scheme. During the 1st phase, the BS directly sends signals to both users using the NOMA protocol. During the 2nd phase, the CU can be a cooperative relay, forwarding signals to the CEU. This two-phase process efficiently utilizes the RIS's reflective capabilities to optimize signal delivery to both users.

	In the first phase, the BS transmits a signal to both users, which can be represented as 
	\begin{align}
		\label{q1} x = \sqrt{\alpha_cP_S}x_c + \sqrt{\alpha_eP_S}x_e. \end{align} 
Here, $P_S$ is the transmit power of the BS, while $x_c$ and $x_e$ are the signals intended for two users, respectively.  Additionally, {we consider the Rayleigh fading environment\footnote{ {The flat Rayleigh fading model is applicable to non-line-of-sight (NLoS) environments with numerous scatterers, such as urban areas, and is widely adopted in practice \cite{Chizhik03, LaiX19}. While RIS-assisted wireless communications typically assume line-of-sight (LoS) connections, LoS links may be unavailable due to strong attenuation from deep fading or shadowing effects.}}}, and $\alpha_c$ and $\alpha_e$ are power allocation factors, satisfying $\alpha_c + \alpha_e = 1$ and $\alpha_c < \alpha_e$. Consequently, the received signals at the CU and CEU are given by
\begin{align}\label{q2}
		y_c=&\left(h_c+\sum_{r=1}^R g_{rc}\eta_{c}e^{j\Phi_{rc}}h_{rc}\right)x+n_c,\\
		\label{q3}y_{e1}=&\left(h_e+\sum_{r=1}^R g_{re}\eta_{e}e^{j\Phi_{re}}h_{re}\right)x+n_{e1},
	\end{align}
	where  $h_c \sim \mathcal{CN}\left(0, \lambda_c\right)$ and $h_e \sim \mathcal{CN}\left(0, \lambda_e\right)$ represent the direct channels from the BS to both the CU and CEU, respectively. Moreover, $h_{rc}\sim \mathcal{CN}\left(0,\lambda_{rc}\right)$, $h_{re}\sim \mathcal{CN}\left(0,\lambda_{re}\right)$, $g_{rc}\sim \mathcal{CN}\left(0,\lambda_{gc}\right)$, and $g_{re}\sim \mathcal{CN}\left(0,\lambda_{ge}\right)$ represent the channel parameters from the BS to the $r$th reflecting element of the $1$st and $2$nd zones, as well as the channel parameters from the $r$th reflecting element of the 1-st and 2-nd zones to the CU and the CEU, respectively.
	Furthermore, $\eta_{c}$ and $\eta_{e}$ represent the {reflection coefficient} of each RIS zone, while $\Phi_{rc}$ and $\Phi_{re}$ denote the reflection coefficient and phase shift from the $r$th reflecting element of the $1$-st and $2$-nd zones to the users. Additionally, $n_c\sim \mathcal{CN}(0,\sigma^2)$ and $n_{e1}\sim \mathcal{CN}(0,\sigma^2)$ are the additive Gaussian noise components at the CU and the CEU, respectively, with $\sigma^2$ representing the noise variance.

After receiving the signal $y_c$, the CU decodes the signal  $x_e$ by regarding the signal $x_c$  as the interference. According to \cite{RenH22}, the optimal $\Phi_{rc}$ that maximizes the instantaneous signal-to-noise-plus interference ratio (SINR)  is given by
	\begin{align}\label{q4}
		\Phi_{rc}=&-\phi_{h_{rc}}-\phi_{g_{rc}},
	\end{align}
where  $\phi_{h_{rc}}$ and $\phi_{g_{rc}}$ are the phases of the channels $h_{rc}$ and $g_{rc}$, respectively.
Besides, the instantaneous SINR of decoding $x_c$ at the CU is given by
	\begin{align}\label{q5} \gamma_{ce}=\frac{\alpha_e\rho_S\left(|h_c|^2+q_c^2\eta_c^2\right)}{\alpha_c\rho_S\left(|h_c|^2+q_c^2\eta_c^2\right)+1},
	\end{align}
	where $\rho_S=\frac{P_S}{\sigma^2}$ is the ratio of the transmit power and the variance of the noise, and $q_c=\sum_{r=1}^R|g_{rc}||h_{rc}|$.  According to \cite{Polyanskiy10}, the instantaneous BLER of decoding $x_e$  at the CU is
	\begin{align}\label{q6}
		\epsilon_{ce}\approx &\Psi(\gamma_{ce},N_e,m),\nonumber\\
		\triangleq & Q\left(\frac{ C(\gamma_{ce})-N_e/m}{\sqrt{V(\gamma_{ce})/m}}\right),
	\end{align}
	where
	\begin{align}\label{q7}
		Q(x)=&\frac{1}{\sqrt{2\pi}}\int_{x}^{\infty}e^{-\frac{t^2}{2}}dt,\\
		\label{q8}C(\gamma)=&\log_2(1+\gamma),\\
		\label{q9}V(\gamma)=&(\log_2 e)^2\cdot(1-(1+\gamma)^{-2}),
	\end{align} 
 $m$  represents the blocklength, while $N_e$  denotes the number of data bits from BS to the CEU. 
	
	According to the SIC decoding scheme, once $x_e$ is successfully decoded by the CU, $x_c$ will be decoded consequently. Hence, the instantaneous SNR of decoding $x_c$ can be written as
	\begin{align}\label{q10}
		\gamma_{cc}=\alpha_c\rho_S\left(|h_c|^2+q_c^2\eta_c^2\right).
	\end{align}
As a result, the instantaneous BLER of decoding $x_c$  at the CU is
	\begin{align}\label{q11}
		\epsilon_{cc}\approx &\Psi(\gamma_{cc},N_c,m),\nonumber\\
		\triangleq & Q\left(\frac{ C(\gamma_{cc})-N_c/m}{\sqrt{V(\gamma_{cc})/m}}\right),
	\end{align}
	where $N_c$ is the number of data bits from BS to the CU.
	
	Thus, the instantaneous BLER when CU decoding $y_c$ can be written as
	\begin{align}\label{q1w}
		\epsilon_c =\epsilon_{ce}+\epsilon_{cc}-\epsilon_{ce}\epsilon_{cc}.
	\end{align}
	
	During the first phase, the CEU receives the signal $y_e$ and decodes the signal  $x_e$, treating the signal $x_c$ as interference.  The optimal $\Phi_{re}$ that maximizes the instantaneous SINR is given by \cite{LiS22}
	\begin{align}\label{q13}
		\Phi_{re}=&-\phi_{h_{re}}-\phi_{g_{re}},
	\end{align}
where  $\phi_{h_{re}}$ and $\phi_{g_{re}}$  denote the phases of the channels $h_{re}$ and $g_{re}$, respectively.
The instantaneous SINR for decoding $x_e$ at the CEU can be expressed as
	\begin{align}\label{q14}		\gamma_{e1}=\frac{\alpha_e\rho_S\left(|h_e|^2+q_e^2\eta_e^2\right)}{\alpha_c\rho_S\left(|h_e|^2+q_e^2\eta_e^2\right)+1},
	\end{align}
	where $q_e=\sum_{r=1}^R|g_{re}||h_{re}|$.  As per \cite{Polyanskiy10}, the BLER for decoding $x_e$  at the CEU is
	\begin{align}\label{q15}
		\epsilon_{e1} \approx &\Psi(\gamma_{e1},N_e,m),\nonumber\\
		\triangleq & Q\left(\frac{ C(\gamma_{e1})-N_e/m}{\sqrt{V(\gamma_{e1})/m}}\right).
	\end{align}
	
	During the 2-nd phase,  the CU acts as a relay, transmitting the signal $x_e$ to the CEU with a transmit power of $P_c$.  The signal received by the CEU can be written as
	\begin{align}\label{q16}
		y_{e2}=\sqrt{P_c}x_e\left(h_{ce}+\sum_{r=1}^R g_{rce}\eta_e e^{j\Phi_{rce}}h_{rce}\right)+n_{e2},
	\end{align}
	where  { $h_{ce}\sim \mathcal{CN}\left(0,\lambda_{ce}\right)$, $g_{rce}\sim \mathcal{CN}\left(0,\lambda_{gce}\right)$, $h_{rce}\sim \mathcal{CN}\left(0,\lambda_{rce}\right)$}, $\Phi_{rce}$ represents the phase shift from the $r$th reflecting element of $2$-nd zones to the CEU. The noise $n_{e2}\sim \mathcal{CN}(0,\sigma^2)$ is the additive Gaussian noise at the CEU. The optimal phase shift $\Phi_{rce}$ maximizing the instantaneous SINR  is given by\cite{RenH22}
	\begin{align}\label{q17}
		\Phi_{rce}=&-\phi_{h_{rce}}-\phi_{g_{rce}},
	\end{align}
	where  $\phi_{h_{rce}}$ and $\phi_{g_{rce}}$ are the phases of the channels $h_{rce}$ and $g_{rce}$, respectively. The the instantaneous SNR for decoding $x_e$ at the CEU is given by
	\begin{align}\label{q18}
		\gamma_{e2}=\rho_C\left(|h_{ce}|^2+q_{ce}^2\eta_e^2\right),
	\end{align}
	with  $\rho_C=\frac{P_C}{\sigma^2}$ representing  the ratio between the transmit power of the CU and the noise variance. Further, $q_{ce}=\sum_{r=1}^R|g_{rce}||h_{rce}|$. As per  \cite{Polyanskiy10}, the instantaneous BLER for decoding $x_e$  at the CEU is
	\begin{align}\label{q19}
		\epsilon_{e2}\approx &\Psi(\gamma_{e2},N_e,m),\nonumber\\
		\triangleq & Q\left(\frac{ C(\gamma_{e2})-N_e/m}{\sqrt{V(\gamma_{e2})/m}}\right).
	\end{align}
To enhance the signal integration at the CEU during both transmission phases, we propose to employ SC and MRC. Using SC, the SINR for decoding $x_e$ is given by
\begin{align}\label{q20}
	\gamma_e=\max\{\gamma_{e1}, \gamma_{e2}\}.
\end{align}
Meanwhile, with MRC, the SINR is
\begin{align}\label{q21}
	\bar{\gamma}_e=\gamma_{e1}+\gamma_{e2}.
\end{align}
Accordingly, the instantaneous BLER for decoding $x_e$ at the CEU is expressed as
\begin{align}\label{q22}
	\epsilon_e=\epsilon_{ce}\Psi(\gamma_{e1}, N_e, m)+(1-\epsilon_{ce})\Psi(\gamma_e,N_e,m)
\end{align}
when using  SC, and as
\begin{align}\label{q23}
	\bar{\epsilon}_e=\epsilon_{ce}\Psi(\gamma_{e1}, N_e, m)+(1-\epsilon_{ce})\Psi(\bar{\gamma}_e,N_e,m)
\end{align}
when employing MRC.

	\section{ Average BLER of the Proposed Systems}
This section provides a detailed theoretical analysis of the average BLER for decoding $x_e$ at the CEU and $x_c$ at the CU.
\subsection{Derivation of $\mathbb{E}[\epsilon_{c}]$}
 The derivation for the average BLER associated with CU decoding
$x_c$ is as follows
	\begin{align}\label{q24}
		\mathbb{E} [\epsilon_c]=&\mathbb{E}[\epsilon_{cc}]+\mathbb{E}[\epsilon_{ce}]-\mathbb{E}[\epsilon_{cc}\epsilon_{ce}].
	\end{align}
In the subsequent subsections, we will systematically derive the expressions for $\mathbb{E}[\epsilon_{cc}]$, $\mathbb{E}[\epsilon_{ce}]$, and $\mathbb{E}[\epsilon_{ce}]$.
\subsubsection{Derivation of $\mathbb{E}[\epsilon_{cc}]$}
To begin, the expression for $\mathbb{E}[\epsilon_{cc}]$ is provided as:
	\begin{align}\label{q25}
		\mathbb{E}\left[\epsilon_{cc}\right]\approx\int_{0}^{\infty}\Psi(\gamma_{cc},N_c,m)f_{\gamma_{cc}}(t)dt.
	\end{align}
To compute the results of $\mathbb{E}\left[\epsilon_{cc}\right]$, we apply a linear approximation for $\Psi(\gamma_{cc},N_c,m)$,  detailed below \cite{YuY18,Makki14}
	\begin{align}\label{q26}
		&\Psi(\gamma_{cc},N_c,m)\nonumber\\&=\left\{
		\begin{array}{ll}
			1, & \gamma_{cc}\leq v_{N_c,m} \\
			\frac{1}{2}-\delta_{N_c,m}\sqrt{n}(\gamma_{cc}-\beta_{N_c,m}), &v_{N_c,m}<\gamma_{cc}<u_{N_c,m} \\
			0, & \gamma_{cc}\geq u_{N_c,m}
		\end{array}
		\right.
	\end{align}
	where
	\begin{align}\label{q27}
		\beta_{N_c,m}=&2^\frac{N_c}{m}-1,\\
		\label{q28}\delta_{N_c,m}=&(2\pi(2^\frac{2N_c}{m}-1))^{-\frac{1}{2}},\\
		\label{q29}v_{N_c,m}=&\beta_{N_c,m}-\frac{1}{2}\delta_{N_c,m}^{-1}m^{-\frac{1}{2}},\\
		\label{q30}u_{N_c,m}=&\beta_{N_c,m}+\frac{1}{2}\delta_{N_c,m}^{-1}m^{-\frac{1}{2}}.
	\end{align}
	
Incorporating  \eqref{q26} into \eqref{q25}, the approximation is provided as follows \cite{YuY18}:
	\begin{equation}\label{q31}	\mathbb{E}\left[\epsilon_{cc}\right]\approx\delta_{N_c,m}\sqrt{m}\int_{v_{N_c,m}}^{u_{N_c,m}}F_{\gamma_{cc}}(t)dt,
	\end{equation}
	where $F_{\gamma_{cc}}(t)$ represents the CDF of $\gamma_{cc}$.
 Given the complexity, a direct calculation of the integral for $F_{\gamma_{cc}}(t)$ is challenging. Therefore, to further elucidate the expression of $\mathbb{E}\left[\epsilon_{cc}\right]$,  we present the subsequent Lemmas addressing $\mathbb{E}\left[\epsilon_{cc}\right]$.
	
	\emph{Lemma 1:}  Let $T=|h_c|^2+q_c^2\eta_c^2$, the CDF of $T$ is given by
	\begin{align}\label{q32}
		F_T(t)=&\frac{\gamma\left(\kappa+1,\frac{\sqrt{t}}{\eta_c b_c}\right)}{\Gamma\left(\kappa+1\right)}-\frac{\sqrt{t}}{2\eta_cb_c^\kappa\Gamma\left(\kappa+1\right)}\exp\left(-\frac{t}{\lambda_c}\right)\nonumber\\
		\cdot&\sum_{u=1}^U\sqrt{1-\xi_u^2}\frac{\pi}{U}\zeta_{uc}^{\kappa}(t)\exp\left(\frac{\eta_c^2\zeta_{uc}^2(t)}{\lambda_c}-\frac{\zeta_{uc}(t)}{b_c}\right).
	\end{align}
	where
	\begin{align}\label{q33}
		\kappa=&\frac{\left(R+1\right)\pi^2-16}{16-\pi^2},\\
		\label{q34}b_c=&\left(\frac{4}{\pi}-\frac{\pi}{4}\right)\sqrt{\lambda_{gc}\lambda_{rc}},
	\end{align}
	and
	\begin{align}\label{q35}
		\Gamma(x)=\int_0^\infty t^{x-1}e^{-t}dt
	\end{align}
	is the gamma function\cite{Gradshteyn07}, and
	\begin{align}\label{q36}
		\gamma\left(\alpha,x\right)=\int_0^x e^{-t} t^{\alpha-1}dt
	\end{align}
	is the lower incomplete gamma function\cite{Gradshteyn07}. Moreover,
	\begin{align}\label{q39}
		\xi_u=&\cos\left(\frac{2u-1}{2U}\pi\right),\\
		\label{q40}\zeta_{uc}(t)=&\frac{\sqrt{t}}{2\eta_c}\left(\xi_u+1\right),
	\end{align}
	where  $U$ denotes an accuracy-complexity tradeoff parameter.

	\emph{Proof:} See Appendix \ref{A1}. \hfill$\blacksquare$
	
	Consequently, we can derive the CDF of $\gamma_{cc}$ in Lemma 2 according to\emph{ Lemma 1}.
	
	\emph{Lemma 2:}  The CDF of $\gamma_{cc}$ is given in Appendix \ref{A2}.

	where
	\begin{align}\label{q377}
		\zeta_{uc}\left(\frac{\omega}{\alpha_c\rho_S}\right)=\frac{\sqrt{\frac{\omega}{\alpha_c\rho_S}}}{2\eta_c}\left(\xi_u+1\right).
	\end{align}
	
	\emph{Proof:} See Appendix \ref{A3}. \hfill$\blacksquare$
	
	When $m$ is sufficiently large, the integral interval $[v_{N_c,m},u_{N_c,m}]$ tends to be small. Therefore, we can use the first-order Riemann integral approximation given by
	\begin{align}\label{q84}
		\int_{v_{N_c,m}}^{u_{N_c,m}}F_{\gamma_{cc}}(t)dt=\left(u_{N_c,m}-v_{N_c,m}\right)F_{\gamma_{cc}}\left(\frac{v_{N_c,m}+u_{N_c,m}}{2}\right).
	\end{align}
By substituting \eqref{q37} into \eqref{q31}, we obtain $\mathbb{E}\left[\epsilon_{cc}\right]$ as given in Appendix \ref{A4}.  
	
\subsubsection{Derivation of $\mathbb{E}\left[\epsilon_{ce}\right]$}	
To derive the expression of $\mathbb{E}\left[\epsilon_{ce}\right]$, we need to derive the CDF of $\gamma_{ce}$ first.

Similar to derive the CDF of $\gamma_{cc}$, we can derive the CDF of $\gamma_{ce}$ in \emph{Lemma 3} according to \emph{Lemma 1}.
	
	\emph{Lemma 3:} The CDF of $\gamma_{ce}$ is given by \eqref{q42}.
	\begin{figure*}[t]
		\centering
		\begin{align}\label{q42}
			F_{\gamma_{ce}}\left(\omega\right)=\left\{
			\begin{array}{lcl}
				1, && \mbox{if}\ \omega \geq\frac{\alpha_e}{\alpha_c},\\
				\frac{\gamma\left(\kappa+1,\frac{\sqrt{\frac{\omega}{\alpha_e\rho_S-\alpha_c\rho_S \omega}}}{\eta_c b_c}\right)}{\Gamma\left(\kappa+1\right)}
				-\frac{\sqrt{\frac{\omega}{\alpha_e\rho_S-\alpha_c\rho_S \omega}}}{2\eta_cb_c^\kappa\Gamma\left(\kappa+1\right)}\exp\left(-\frac{\omega}{\lambda_c\left(\alpha_e\rho_S-\alpha_c\rho_S \omega\right)}\right)\\
				\cdot\sum_{u=1}^U\sqrt{1-\xi_u^2}\frac{\pi}{U}\zeta_{uc}^{\kappa}\left(\frac{\omega}{\alpha_e\rho_S-\alpha_c\rho_S \omega}\right)\exp\left(\frac{\eta_c^2\zeta_{uc}^2\left(\frac{\omega}{\alpha_e\rho_S-\alpha_c\rho_S \omega}\right)}{\lambda_c}-\frac{\zeta_{uc}\left(\frac{\omega}{\alpha_e\rho_S-\alpha_c\rho_S \omega}\right)}{b_c}\right), && \mbox{otherwise}.
			\end{array}\right.
		\end{align}\hrule
	\vspace{1ex}
	\end{figure*}
	
	\emph{Proof:} See Appendix \ref{A5}.  \hfill$\blacksquare$

	We use a linear approximation of function to compute the results of $\mathbb{E}\left[\epsilon_{ce}\right]$,  which is given by \cite{YuY18,Makki14}
	\begin{align}\label{q43}
		&\Psi(\gamma_{ce},N_e,m)\nonumber\\&=\left\{
		\begin{array}{ll}
			1, & \gamma_{ce}\leq v_{N_e,m} \\
			\frac{1}{2}-\delta_{N_e,m}\sqrt{n}(\gamma_{ce}-\beta_{N_e,m}), &v_{N_e,m}<\gamma_{ce}<u_{N_e,m} \\
			0, & \gamma_{ce}\geq u_{N_e,m}
		\end{array}
		\right.
	\end{align}
	where
	\begin{align}\label{q44}
		\beta_{N_e,m}=&2^\frac{N_e}{m}-1,\\
		\label{q45}\delta_{N_e,m}=&(2\pi(2^\frac{2N_e}{m}-1))^{-\frac{1}{2}},\\
		\label{q46}v_{N_e,m}=&\beta_{N_e,m}-\frac{1}{2}\delta_{N_e,m}^{-1}m^{-\frac{1}{2}},\\
		\label{q47}u_{N_e,m}=&\beta_{N_e,m}+\frac{1}{2}\delta_{N_e,m}^{-1}m^{-\frac{1}{2}}.
	\end{align}
	Similar with the derivation of $\mathbb{E}\left[\epsilon_{cc}\right]$, we have \cite{YuY18}
	\begin{equation}\label{q48}
		\mathbb{E}\left[\epsilon_{ce}\right]\approx\delta_{N_e,m}\sqrt{m}\int_{v_{N_e,m}}^{u_{N_e,m}}F_{\gamma_{ce}}(t)dt,
	\end{equation}
	Thus, when $t>\frac{\alpha_e}{\alpha_c}$,
	\begin{equation}\label{q49}
		\mathbb{E}\left[\epsilon_{ce}\right]= 1,
	\end{equation}
	otherwise, the first order Riemann integral approximation, we obtain $\mathbb{E}\left[\epsilon_{ce}\right]$ is given in Appendix \ref{A4}. 
\subsubsection{Derivation of $\mathbb{E}\left[\epsilon_{cc}\epsilon_{ce}\right]$}	
	Based on \textbf{Proposition 1} from \cite{LaiX19}, we can express
\begin{align}\label{q53}
	\mathbb{E}[\epsilon_{cc}\epsilon_{ce}] \approx \min\{\mathbb{E}[\epsilon_{cc}],\mathbb{E}[\epsilon_{ce}]\}.
\end{align}
From this, it follows that the analytical average BLER for decoding $x_c$ at the CU can be articulated as
\begin{align}\label{q54}
	\mathbb{E}[\epsilon_{c}] \geq \max\{\mathbb{E}[\epsilon_{cc}],\mathbb{E}[\epsilon_{ce}]\}.
\end{align}
By incorporating \eqref{q38} and \eqref{q53} into \eqref{q54}, we arrive at Eqn. \eqref{q55} in Appendix \ref{A4}. 
	where
	\begin{align}
		\Theta\left(x\right)=\left\{
		\begin{array}{ll}
			1,  &  x>0\\
			0,  &  x\leq 0
		\end{array}
		\right.
	\end{align}
	is the unit step function.
	\subsection{Derivation of $\mathbb{E}[\epsilon_{e}]$}
In this subsection, we elucidate the derivation of $\mathbb{E}[\epsilon_{e}]$ in the context of employing the SC scheme and MRC scheme.
\subsubsection{SC scheme}
Within the context of the SC scheme, the analytical average BLER for decoding $x_e$ at the edge user is represented as
	\begin{align}\label{q56}
		\mathbb{E} [\epsilon_e]=&\mathbb{E}[\epsilon_{ce}]\mathbb{E}[\Psi(\gamma_{e1}, N_e, m)]\nonumber\\
		&+(1-\mathbb{E}[\epsilon_{ce})])\mathbb{E}[\Psi(\gamma_e,N_e,m)].
	\end{align}
	Building upon this expression, it becomes imperative to elucidate the precise formulations for $\mathbb{E}[\Psi(\gamma_{e1}, N_e, m)]$ and $\mathbb{E}[\Psi(\gamma_e,N_e,m)]$. To address this, we introduce the subsequent Lemmas.

	\emph{Lemma 4:} Let $Z=|h_e|^2+q_e^2\eta_e^2$, the CDF of $Z$ are given by
	\begin{align}\label{q57}
		F_Z(z)=&\frac{\gamma\left(\kappa+1,\frac{\sqrt{z}}{\eta_e b_e}\right)}{\Gamma\left(\kappa+1\right)}-\frac{\sqrt{z}}{2\eta_eb_e^\kappa\Gamma\left(\kappa+1\right)}\exp\left(-\frac{z}{\lambda_e}\right)\nonumber\\
		\cdot&\sum_{u=1}^U\sqrt{1-\xi_u^2}\frac{\pi}{U}\zeta_{ue}^{\kappa}(z)\exp\left(\frac{\eta_e^2\zeta_{ue}^2(z)}{\lambda_e}-\frac{\zeta_{ue}(z)}{b_e}\right),
	\end{align}
	where
	\begin{align}
		b_e=&\left(\frac{4}{\pi}-\frac{\pi}{4}\right)\sqrt{\lambda_{ge}\lambda_{re}},\\
		\zeta_{ue}(z)=&\frac{\sqrt{z}}{2\eta_e}\left(\xi_u+1\right).
	\end{align}
	
	\emph{Proof:} The proof is similar to Appendix A, and hence is omitted for simplicity.   \hfill$\blacksquare$

Moreover, the CDF of $\gamma_{e1}$ is given by
	\begin{align}\label{q58}
		F_{\gamma_{e1}}(\tau)=&\Pr\left(\gamma_{e1}\leq \tau\right),
	\end{align}
	where $\gamma_{e1}$ can be written as
	\begin{align}\label{q59}
		\gamma_{e1}=\frac{\alpha_e\rho_S Z}{\alpha_c\rho_S Z+\eta_e^2+1},
	\end{align}
	and substituting \eqref{q59} into \eqref{q58} yields:
	\begin{align}\label{q60}
		F_{\gamma_{e1}}(\tau)=&\Pr\left(Z\leq \frac{\tau}{\alpha_e\rho_S-\alpha_c\rho_S\tau}\right),
	\end{align}
	Drawing parallels with $F_{\gamma_{ce}}$, the expression for $F_{\gamma_{e1}}(\tau)$  can be found in \eqref{q62},
	\begin{figure*}[t]
		\centering
		\begin{align}\label{q62}
			F_{\gamma_{e1}}\left(\tau\right)=\left\{
			\begin{array}{lcl}
				1, && \mbox{if}\ \tau \geq\frac{\alpha_e}{\alpha_c},\\
				\frac{\gamma\left(\kappa+1,\frac{\sqrt{\frac{\tau}{\alpha_e\rho_S-\alpha_c\rho_S\tau}}}{\eta_e b_e}\right)}{\Gamma\left(a_e+1\right)}-\frac{\sqrt{\frac{\tau}{\alpha_e\rho_S-\alpha_c\rho_S\tau}}}{2\eta_eb_e^\kappa\Gamma\left(\kappa+1\right)}\exp\left(-\frac{\tau}{\lambda_e\left(\alpha_e\rho_S-\alpha_c\rho_S\tau\right)}\right)\\
				\cdot\sum_{u=1}^U\sqrt{1-\xi_u^2}\frac{\pi}{U}\zeta_{ue}^{\kappa}\left(\frac{\tau}{\alpha_e\rho_S-\alpha_c\rho_S\tau}\right)\exp\left(\frac{\eta_e^2\zeta_{ue}^2\left(\frac{\tau}{\alpha_e\rho_S-\alpha_c\rho_S\tau}\right)}{\lambda_e}-\frac{\zeta_{ue}\left(\frac{\tau}{\alpha_e\rho_S-\alpha_c\rho_S\tau}\right)}{b_e}\right), && \mbox{otherwise}.
			\end{array}\right.
		\end{align}\hrule
	\end{figure*}
	where
	\begin{align}
		\zeta_{ue}\left(\frac{\tau}{\alpha_e\rho_S-\alpha_c\rho_S\tau}\right)=\frac{\sqrt{\frac{\tau}{\alpha_e\rho_S-\alpha_c\rho_S\tau}}}{2\eta_e}\left(\xi_u+1\right).
	\end{align}
	
	Similar with the derivation of $\mathbb{E}\left[\epsilon_{ce},N_e,m)\right]$, we have $\mathbb{E}\left[\Psi(\gamma_{e1},N_e,m)\right]$ as given by \eqref{q65},
	\begin{figure*}[hb]
\hrule 
		\centering
		\begin{align}\label{q65}
			&\mathbb{E}\left[\Psi(\gamma_{e1},N_e,m)\right]\nonumber\\
			\approx& \left(\Theta\left(\beta_{N_e,m}-\frac{\alpha_e}{\alpha_c}\right)+\Theta\left(\frac{\alpha_e}{\alpha_c}-\beta_{N_e,m}\right)\frac{\gamma\left(\kappa+1,\frac{\sqrt{\frac{\beta_{N_e,m}}{\alpha_e\rho_S-\alpha_c\rho_S\beta_{N_e,m}}}}{\eta_e b_e}\right)}{\Gamma\left(a_e+1\right)}-\frac{\sqrt{\frac{\beta_{N_e,m}}{\alpha_e\rho_S-\alpha_c\rho_S\beta_{N_e,m}}}}{2\eta_eb_e^\kappa\Gamma\left(\kappa+1\right)}\exp\left(-\frac{\beta_{N_e,m}}{\lambda_e\left(\alpha_e\rho_S-\alpha_c\rho_S\beta_{N_e,m}\right)}\right)\right.\nonumber\\
			\cdot&\left.\sum_{u=1}^U\sqrt{1-\xi_u^2}\frac{\pi}{U}\zeta_{ue}^{\kappa}\left(\frac{\beta_{N_e,m}}{\alpha_e\rho_S-\alpha_c\rho_S\beta_{N_e,m}}\right)\exp\left(\frac{\eta_e^2\zeta_{ue}^2\frac{\beta_{N_e,m}}{\alpha_e\rho_S-\alpha_c\rho_S\beta_{N_e,m}}}{\lambda_e}-\frac{\zeta_{ue}\frac{\beta_{N_e,m}}{\alpha_e\rho_S-\alpha_c\rho_S\beta_{N_e,m}}}{b_e}\right)\right)
		\end{align} 
	\end{figure*}
	where
	\begin{align}
		\zeta_{ue}\left(\frac{\beta_{N_e,m}}{\alpha_e\rho_S-\alpha_c\rho_S\beta_{N_e,m}}\right)=\frac{\sqrt{\frac{\beta_{N_e,m}}{\alpha_e\rho_S-\alpha_c\rho_S\beta_{N_e,m}}}}{2\eta_e}\left(\xi_u+1\right),
	\end{align}
	and $\Psi(\gamma_{e1},N_e,m)$ is using the linear approximate, is similar to $\Psi(\gamma_{ce},N_e,m)$, and hence is omitted for simplicity.

\begin{figure*}[t] 
	\centering
	\begin{align}\label{q74}
		\mathbb{E}\left[\Psi(\gamma_{e2},N_e,m)\right]\approx&\frac{\gamma\left(\kappa+1,\frac{\sqrt{\frac{\beta_{N_e,m}}{\rho_C}}}{\eta_e b_e}\right)}{\Gamma\left(\kappa+1\right)}-\frac{\sqrt{\frac{\beta_{N_e,m}}{\rho_C}}}{2\eta_eb_{ce}^\kappa\Gamma\left(\kappa+1\right)}\exp\left(-\frac{\beta_{N_e,m}}{\rho_C\lambda_e}\right)\nonumber\\
		\cdot&\sum_{u=1}^U\sqrt{1-\xi_u^2}\frac{\pi}{U}\zeta_{ue}^{\kappa}\left(\frac{\beta_{N_e,m}}{\rho_C}\right)\exp\left(\frac{\eta_e^2\zeta_{ue}^2\left(\frac{\beta_{N_e,m}}{\rho_C}\right)}{\lambda_e}-\frac{\zeta_{ue}\left(\frac{\beta_{N_e,m}}{\rho_C}\right)}{b_{ce}}\right),
	\end{align}
	\hrule
\end{figure*}
where
\begin{align}
	\zeta_{ue}\left(\frac{\beta_{N_e,m}}{\rho_C}\right)=\frac{\sqrt{\frac{\beta_{N_e,m}}{\rho_C}}}{2\eta_e}\left(\xi_u+1\right).
\end{align}
and $\Psi(\gamma_{e2},N_e,m)$ is using the linear approximate, is similar to $\Psi(\gamma_{ce},N_e,m)$, and hence is omitted for simplicity.
Therefore, we have $\mathbb{E}\left[\Psi(\gamma_{e},N_e,m)\right]$ is given in Appendix \ref{A4}.  
	
	According to the Proposition 2 in \cite{LaiX19}, we have
	\begin{align}\label{q66}
		\mathbb{E}\left[\Psi(\gamma_{e},N_e,m)\right]\approx \mathbb{E}\left[\Psi(\gamma_{e1},N_e,m)\right]\mathbb{E}\left[\Psi(\gamma_{e2},N_e,m)\right],
	\end{align}
	where $\mathbb{E}\left[\Psi(\gamma_{e1},N_e,m)\right]$ is given by \eqref{q65}, and
	\begin{align}\label{q67}
		\mathbb{E}\left[\Psi(\gamma_{e2},N_e,m)\right]=\int_0^\infty \Psi(\gamma_{e2},N_e,m)f_{\gamma_{e2}}(x)dx,
	\end{align}
	where $\Psi(\gamma_{e2},N_e,m)$ is using the linear approximate and is similar to $\Psi(\gamma_{ce},N_e,m)$, which is omitted for simplicity, thus,
	\begin{align}\label{q68}
		\mathbb{E}\left[\Psi(\gamma_{e2},N_e,m)\right]\approx \delta_{N_e,m}\sqrt{m}\int_{v_{N_e,m}}^{u_{N_e,m}}F_{\gamma_{e2}}(t)dt,
	\end{align}
	where $F_{\gamma_{e2}}(t)$ is the CDF of $\gamma_{e2}$, and we obtain the Lemma 5.
	
	\emph{Lemma 5:} Let $W=|h_{ce}|^2+q_{ce}^2\eta_e^2$, the CDF of $W$ is
	\begin{align}\label{q69}
		F_W(w)=&\frac{\gamma\left(\kappa+1,\frac{\sqrt{w}}{\eta_e b_{ce}}\right)}{\Gamma\left(\kappa+1\right)}-\frac{\sqrt{w}}{2\eta_eb_{ce}^\kappa\Gamma\left(\kappa+1\right)}\exp\left(-\frac{w}{\lambda_e}\right)\nonumber\\
		\cdot&\sum_{u=1}^U\sqrt{1-\xi_u^2}\frac{\pi}{U}\zeta_{ue}^{\kappa}(w)\exp\left(\frac{\eta_e^2\zeta_{ue}^2(w)}{\lambda_e}-\frac{\zeta_{ue}(w)}{b_{ce}}\right),
	\end{align}
	where
	\begin{align}
		b_{ce}=&\left(\frac{4}{\pi}-\frac{\pi}{4}\right)\sqrt{\lambda_{gce}\lambda_{rce}}\\
		\zeta_{ue}(w)=&\frac{\sqrt{w}}{2\eta_e}\left(\xi_u+1\right).
	\end{align}
	
	\emph{Proof:} The proof is similar to Appendix A, and hence is omitted for simplicity.

	Thus, the CDF of $\gamma_{e2}$ is given by
	\begin{align}\label{q70}
		F_{\gamma_{e2}}(\chi)=&\Pr\left(\gamma_{e2}\leq \chi\right),
	\end{align}
	where $\gamma_{e2}$ can be written as
	\begin{align}\label{q71}
		\gamma_{e2}=\frac{\rho_C W}{\eta_e^2+1}.
	\end{align}
	By substituting \eqref{q71} into \eqref{q70}, we have
	\begin{align}\label{q72}
		F_{\gamma_{e2}}(\chi)=&\Pr\left(w\leq \frac{\chi}{\rho_C}\right).
	\end{align}
	Add \eqref{q72} into \eqref{q69}, we obtain \eqref{q73},
	\begin{align}\label{q73}
		F_{\gamma_{e2}}(\chi)=&F_W\left(\frac{\chi}{\rho_C}\right)\nonumber\\
		=&\frac{\gamma\left(\kappa+1,\frac{\sqrt{\frac{\chi}{\rho_C}}}{\eta_e b_e}\right)}{\Gamma\left(\kappa+1\right)}\nonumber\\
		&-\frac{\sqrt{\frac{\chi}{\rho_C}}}{2\eta_eb_{ce}^\kappa\Gamma\left(\kappa+1\right)}\exp\left(-\frac{\chi}{\rho_C\lambda_e}\right)\nonumber\\
		\cdot&\sum_{u=1}^U\sqrt{1-\xi_u^2}\frac{\pi}{U}\zeta_{ue}^{\kappa}\left(\frac{\chi}{\rho_C}\right)\nonumber\\
		&\cdot\exp\left(\frac{\eta_e^2\zeta_{ue}^2\left(\frac{\chi}{\rho_C}\right)}{\lambda_e}-\frac{\zeta_{ue}\left(\frac{\chi}{\rho_C}\right)}{b_{ce}}\right),
	\end{align} 
where
\begin{align}
	\zeta_{ue}\left(\frac{\chi}{\rho_C}\right)=\frac{\sqrt{\frac{\chi}{\rho_C}}}{2\eta_e}\left(\xi_u+1\right).
\end{align}
Furthermore,  \eqref{q74} is obtained by the first order Riemann integral approximation.
	
	\subsubsection{MRC  scheme}
Under the MRC scheme, the average BLER for decoding $x_e$ at the CEU is expressed as
	\begin{align}\label{q78}
		\mathbb{E} [\bar{\epsilon}_e]=&\mathbb{E}[\epsilon_{ce}]\mathbb{E}[\Psi(\gamma_{e1}, N_e, m)]\nonumber\\
		&+(1-\mathbb{E}[\epsilon_{ce})])\mathbb{E}[\Psi(\bar{\gamma}_e,N_e,m)].
	\end{align}
	It is noted that \cite{LaiX19,Makki14}
	\begin{align}\label{q79}
		\Psi(\bar{\gamma}_e,N_e,m)\geq \Psi(2\gamma_{e1},N_e,m)\Psi(2\gamma_{e2},N_e,m).
	\end{align}
	We can obtain
	\begin{align}\label{q80}
		\mathbb{E}[\Psi(\bar{\gamma}_e,N_e,m)]\geq \mathbb{E}[\Psi(2\gamma_{e1},N_e,m)]\mathbb{E}[\Psi(2\gamma_{e2},N_e,m)].
	\end{align}
	And $\mathbb{E}[\Psi(2\gamma_{e1},N_e,m)]$ is given in Appendix \ref{A4}, where
	\begin{align}
		\zeta_{ue}\left(\frac{\beta_{N_e,m}}{2\left(\alpha_e\rho_S-\alpha_c\rho_S\beta_{N_e,m}\right)}\right)=&\frac{\sqrt{\frac{\beta_{N_e,m}}{2\left(\alpha_e\rho_S-\alpha_c\rho_S\beta_{N_e,m}\right)}}}{2\eta_e}\nonumber\\
		\cdot&\left(\xi_u+1\right).
	\end{align}
	And $\mathbb{E}[\Psi(2\gamma_{e2},N_e,m)]$ is given by 
		\begin{align}\label{q82}
		&\mathbb{E}\left[2\Psi(\gamma_{e2},N_e,m)\right]\nonumber\\
		&\approx\frac{\gamma\left(\kappa+1,\frac{\sqrt{\frac{\beta_{ue}}{2\rho_C}}}{\eta_e b_e}\right)}{\Gamma\left(\kappa+1\right)}-\frac{\sqrt{\frac{\beta_{ue}}{2\rho_C}}}{2\eta_eb_{ce}^\kappa\Gamma\left(\kappa+1\right)}\exp\left(-\frac{\beta_{ue}}{2\rho_C\lambda_e}\right)\nonumber\\
		&\cdot\sum_{u=1}^U\sqrt{1-\xi_u^2}\frac{\pi}{U}\zeta_{ue}^{\kappa}\left(\frac{\beta_{ue}}{2\rho_C}\right)\exp\left(\frac{\eta_e^2\zeta_{ue}^2\left(\frac{\beta_{ue}}{2\rho_C}\right)}{\lambda_e}-\frac{\zeta_{ue}\left(\frac{\beta_{ue}}{2\rho_C}\right)}{b_{ce}}\right),
	\end{align}
	 where
	\begin{align}
		\zeta_{ue}\left(\frac{\beta_{ue}}{2\rho_C}\right)=&\frac{\sqrt{\frac{\beta_{ue}}{2\rho_C}}}{2\eta_e}\left(\xi_u+1\right).
	\end{align} 
	The proof is similarity with  $\mathbb{E}\left[\Psi(\gamma_{e2},N_e,m)\right]$ and $\mathbb{E}\left[\Psi(\gamma_{e2},N_e,m)\right]$, and hence is omitted for simplicity. By substiuting \eqref{q50},  \eqref{q65},  \eqref{q81} and  \eqref{q82} into  \eqref{q78}, we can obtain $\mathbb{E} [\bar{\epsilon}_e]$.

{\section{Diversity Order}}
{At high SNR, the BLER slope can be used to determine the diversity order. The series form of the lower incomplete gamma function is
	\begin{align}
		\gamma(a,b)=\Gamma(a)\exp(-b)b^{a}\sum_{k=0}^\infty\frac{b^k}{\Gamma(a+k+1)}.
	\end{align}
	Thus, the diversity order can be calculated as \cite{XiaC24},
	\begin{align}\label{q89}
		D=-\lim_{\rho_s \rightarrow \infty}\frac{\log_2 \mathbb{E}[\epsilon]}{\log_2 \rho_s}.
	\end{align}	
Substituting  \eqref{q55} into \eqref{q89}, we have the diversity order at the CU as follows:
	\begin{align}
		D_c\geq \frac{\kappa+1}{2}.
	\end{align}
	
Substituting \eqref{q56} and \eqref{q78} into \eqref{q89}, we have the diversity order at the CEU as follows
	\begin{align}
		D_e^\text{SC}= \frac{\kappa+1}{2},
	\end{align}
	for SC,
	\begin{align}
		D_e^\text{MRC}= \frac{(\kappa+1)^2}{4},
	\end{align}
	for MRC. Accordingly, we have the following remark.
}
\\\textbf{ {Remark 1}}: 
\textit{In the context of the RIS-assisted downlink cooperative NOMA systems, the use of MRC  leads to {a higher diversity order than SC}, as indicated by the formulas (\(\frac{(\kappa+1)^2}{4}\) for MRC versus \(\frac{\kappa+1}{2}\) for SC). This higher diversity order in MRC implies a more reliable performance of the proposed system. By leveraging multiple signal paths through RIS with phase shift optimization, MRC significantly enhances the received signal strength and reduces the probability of error as the SNR increases, particularly benefiting the CEU  in challenging reception conditions.  }

\section{Numerical Results}

This section presents numerical results to evaluate the accuracy of the analytical BLERs. In our presented simulations, we assume  {that the CU is close to the BS and RIS and has better channel conditions, and the CEU is far away from the BS and RIS and has poor channel conditions, so the parameters are as follows:} $\lambda_c=\lambda_{rc}=\lambda_{re}=\lambda_{ce}=\lambda_{rce}=1$, $\lambda_e=\lambda_{ge}=0.3$, $\lambda_{gc}=\lambda_{gce}=0.8$ \cite{LiS22}, and $P_S=10 P_C$. The power allocation coefficient is $\alpha_c=0.1$,  $\alpha_e=0.9$\cite{LaiX19}. The  blocklength is $m=100$with the number of data bits for the CU and CEU being $N_c=300$ bits and $N_e=100$ bits \cite{LaiX19}, respectively.

\begin{figure*}[t] 
	\vspace{-3mm}
	\begin{minipage}{0.32\linewidth}
		\centering
		\includegraphics[width=0.98\textwidth]{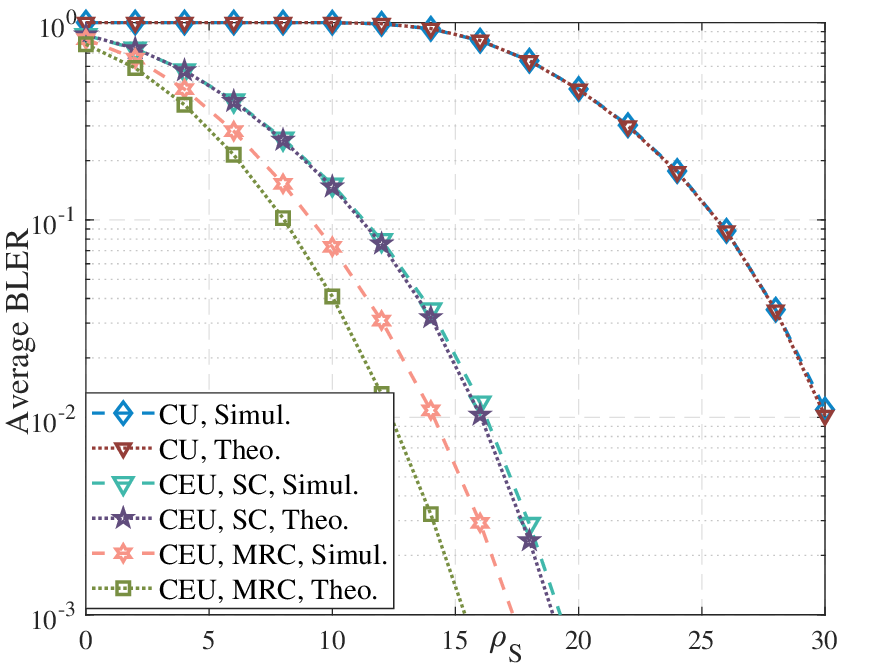}
		\caption{Average BLER versus $\rho_S$, where $m = 100$, $N_c = 300$ bits, $N_e = 100$ bits, $R=8$, $\alpha_c = 0.1$, and $\alpha_e = 0.9$.}
		\label{fs2} 
	\end{minipage}
	\hfill
	\begin{minipage}{0.32\linewidth} 
		\centering
		\includegraphics[width= 1.05\textwidth]{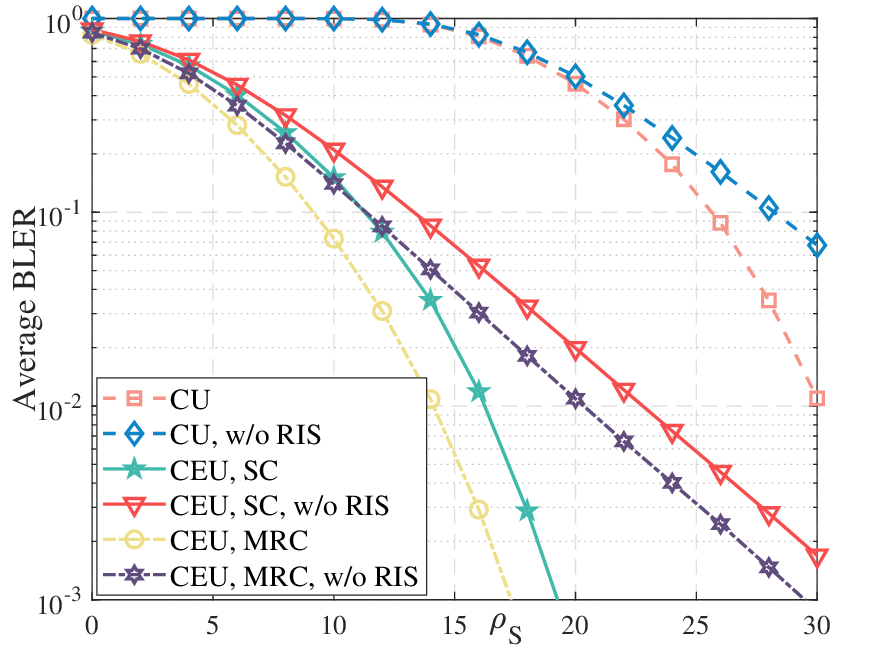}
		\caption{Average BLER versus $\rho_S$ considering the scenarios that downlink NOMA systems with RIS and without RIS.}
		\label{ris} 
	\end{minipage}
	\hfill
	\begin{minipage}{0.32\linewidth} 
		\centering
		\includegraphics[width=0.98\textwidth]{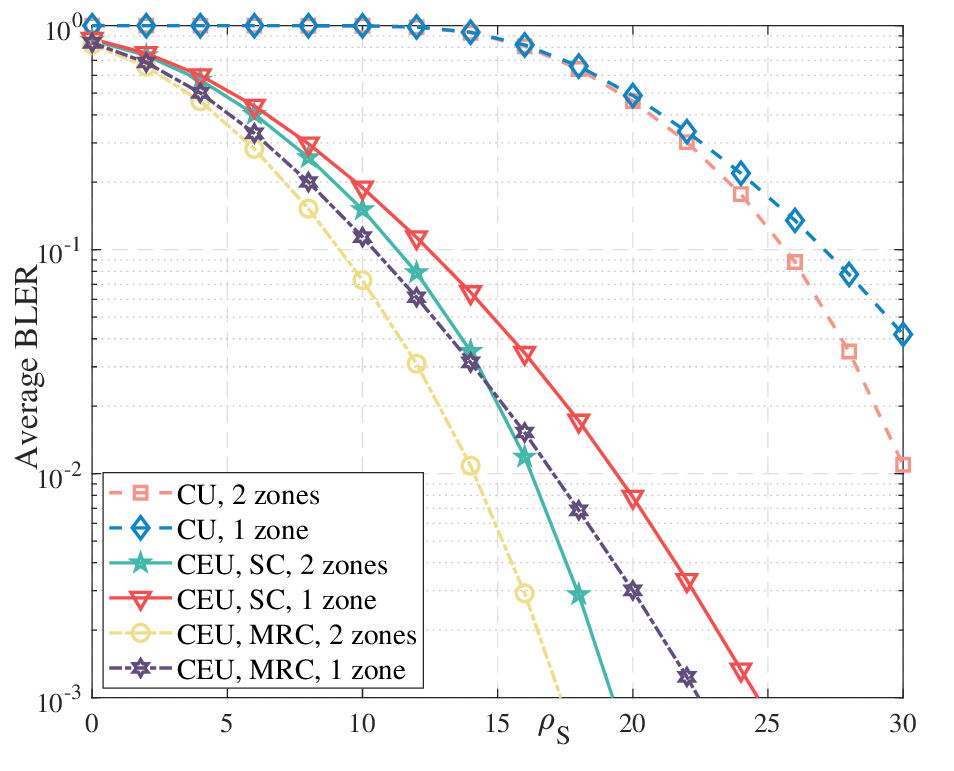}
		\caption{Average BLER versus $\rho_S$ between using an RIS divided into two zones and an RIS with a single zone.}
		\label{zone}
	\end{minipage} 
\end{figure*}

In Fig. \ref{fs2}, we depict the average BLER of the CU and the CEU versus $\rho_S$ in RIS-aided cooperative NOMA system, where $R=8$. In Fig.2, the simulated average BLER, the theoretical analysis one obtained in Section III for the cenreal user and the CEU are denoted by ``CU, Simul.", ``CU, Theo.", and ``EU, SC, Simul.", and  ``EU, SC, Theo.", and ``EU, MRC, Simul.", and ``EU, MRC, Theo.", respectively. And the simulated average BLER without RIS for the cenreal user and the CEU are denoted by ``CU, WO, Simul." and ``EU, WO, Simul.", respectively. 

Furthermore, as shown in Fig. \ref{fs2}, the theoretical analysis of average BLERs for both CU and CEU under the SC scheme is almost identical to the simulated results based on our approach, while the results for MRC represent only the lower bound of the average BLER for the CEU. This discrepancy for MRC can be attributed to the challenge of achieving accurate phase alignment and weighting of signals reflected from different paths in a practical RIS-aided NOMA system. 
\\\textbf{{Remark 2}}: 
\textit{The performance discrepancy between the simulated and theoretical values for MRC in RIS-aided NOMA systems can be attributed to difficulties in accurately aligning and weighting signals from multiple reflected paths. As the number of reflected paths increases, phase misalignment errors accumulate, diminishing MRC's effectiveness. }

In Fig. \ref{ris}, we illustrate the average BLER of the CU and CEU as a function of $\rho_S$ in a downlink cooperative NOMA system, both with and without RIS. The results demonstrate that the performance of the RIS-aided cooperative NOMA system is significantly better than that of the system without RIS, confirming the positive impact of RIS on overall system performance. Additionally, Fig. \ref{ris} shows that in the proposed RIS-aided cooperative NOMA network, the MRC scheme outperforms the SC scheme. Notably, the influence of RIS on the CEU is more pronounced, effectively enhancing the wireless communication performance of the CEU. Furthermore, MRC yields a more significant improvement compared to SC, which is due to the fact that MRC leverages the combination of signals from multiple reflected paths, fully utilizing the diversity gains provided by RIS, while SC only selects the strongest path, limiting its ability to exploit the full potential of the RIS.

Fig. \ref{zone} compares the average BLER performance between using an RIS divided into two zones and an RIS with a single zone in a downlink cooperative NOMA system.  {We assume the RIS reflection angle for any single-zone user is uniformly distributed over $[0, 2\pi]$  \cite{VuT22}, i.e., it does not align any user.} From Fig. \ref{zone}, it is evident that when the RIS is segmented into two zones, both the CU and CEU experience lower average BLER compared to the single-zone RIS case. The improvement is especially significant for the CEU, indicating that dividing the RIS into two zones significantly enhances the CEU's communication performance. This suggests that to effectively boost the communication quality for the CEU, it is preferable to divide the RIS into two zones. Dividing the RIS into two zones allows each zone to focus on optimizing the reflection for a specific user (CU or CEU). In this configuration, the RIS can more effectively direct signals to their intended destinations with tailored phase shifts, improving the overall signal strength and reducing interference. In contrast, a single-zone RIS must balance the signal reflection for both users, leading to suboptimal phase shifts and less efficient signal delivery. By separating the RIS into two zones, the system achieves better isolation and more precise control of the reflected signals, resulting in lower average BLER, particularly for the CEU, which benefits more from the tailored reflections. Hence, we have the following remark:
\\\textbf{{Remark 3}}: 
\textit{Dividing the RIS into two zones significantly reduces average BLER, especially for the CEU, by enabling more precise signal reflection and interference reduction through user-specific phase reconfiguration.}

\begin{figure}[t]
	\centering 
	\includegraphics[width=3.2in]{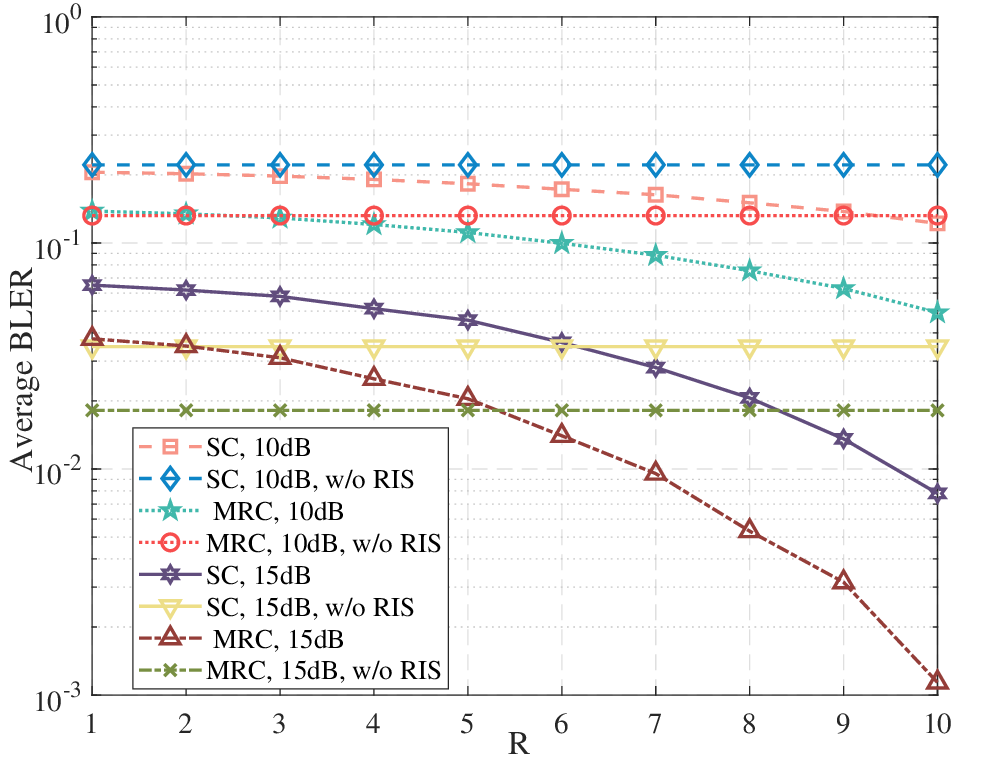}
	\caption{Average BLER versus $R$, where $m = 100$, $N_c = 300$ bits, $N_e = 100$ bits, $\rho_S=10$ dB or $\rho_S=15$ dB, $\alpha_c = 0.1$, and $\alpha_e = 0.9$.}\label{Fig2}
\end{figure}

In Fig. \ref{Fig2}, we present the average BLER  of the CEU  {as a function of the number $R$ of reflection elements}, where $\rho_S=10$ dB and $\rho_S=15$ dB, respectively. From Fig. \ref{Fig2}, it is evident that as the number of reflection elements  $R$ increases, the average BLER decreases. This indicates that the performance improves with more RIS elements, which is expected as additional elements enhance the reflected signal strength, leading to better communication quality. {Nevertheless, RIS hardware becomes more complex and costly with more RIS elements.} In the case of higher SNR (\( 15  \text{ dB} \)), the performance gain (i.e., reduction in BLER) is more pronounced compared to the 10 dB case. This indicates that under higher SNR conditions, the effect of increasing \(R\) becomes more significant. The system benefits more from RIS in high SNR conditions. This is due to the fact that  the channel quality is already favorable, and RIS further amplifies these improvements by optimizing the reflection paths.  Across both SNR scenarios, the MRC scheme consistently outperforms the SC scheme, as expected. MRC utilizes the full potential of all diversity branches by coherently combining them, resulting in better signal quality and lower BLER. In contrast, SC only selects the strongest branch, which leads to less efficient performance. Furthermore, the gap between MRC and SC widens at higher SNRs. Under the 15 dB condition, MRC shows significantly lower BLER than SC, indicating that MRC is better suited to exploit the benefits of high SNR and large RIS setups. This highlights the importance of using advanced combining techniques, especially in the high SNR regime.  Hence, we have the following remark:
\\\textbf{{Remark 4}}: 
\textit{The incorporation of RIS significantly enhances the performance of both SC and MRC, with MRC benefiting more due to its ability to better utilize the additional signal power and diversity. At higher SNR (15 dB), the combined effects of RIS and MRC lead to substantial reliability improvements. However, the performance gains diminish as the number of RIS elements \(R\) increases, especially in the cases with favorable channel conditions.}

\begin{figure}[t]
	\centering 
	\includegraphics[width=3.0in]{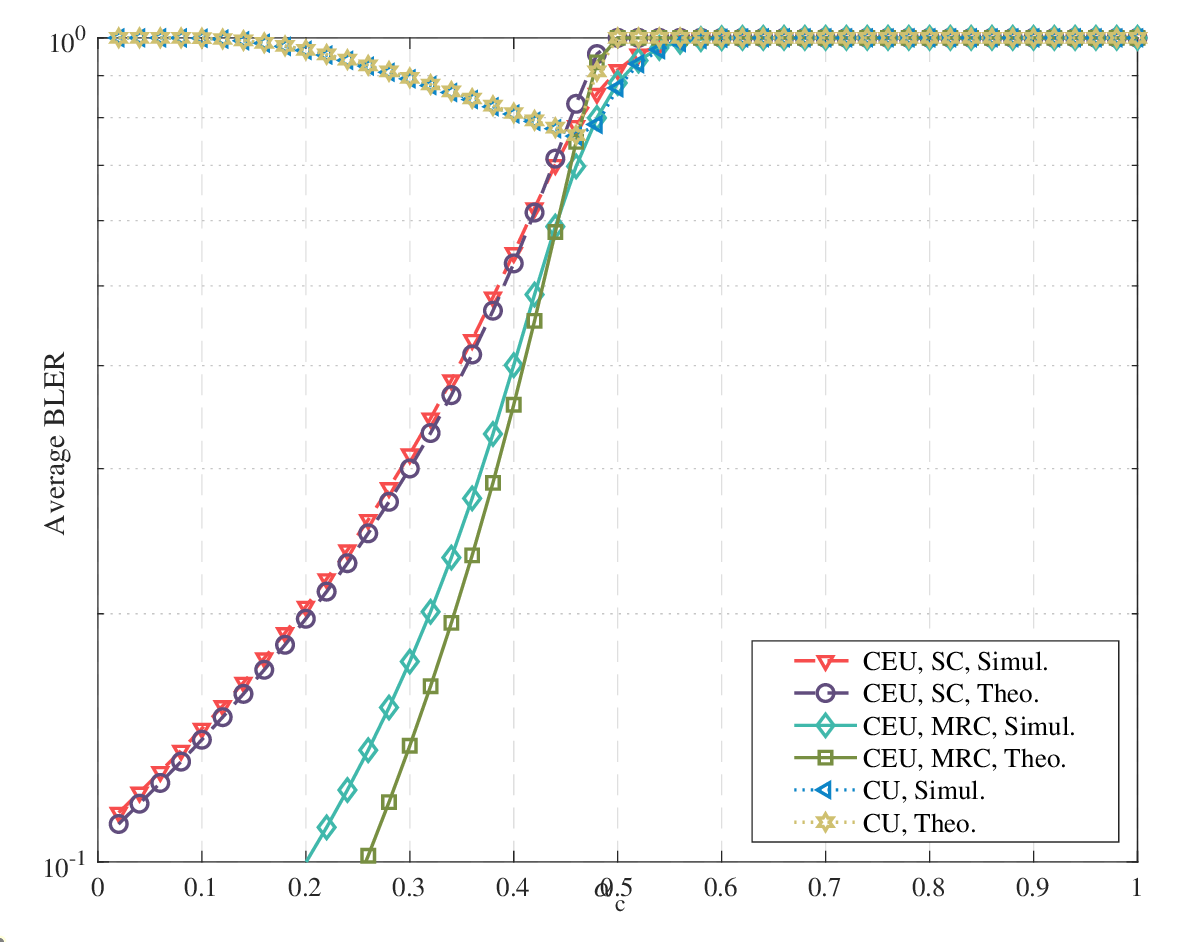}
	\caption{Average BLER versus $\alpha_c$, where $m = 100$, $N_c = 300$ bits, $N_e = 100$ bits, $\rho_S=10$ dB and $R=8$.}\label{Fig3}
\end{figure}

In Fig. \ref{Fig3}, the graph plots the average BLER for the CU and CEU as a function of \(\alpha_c\). For the CU, the average BLER gradually decreases as \(\alpha_c\) increases from 0 to about 0.43, indicating improved performance as more power is allocated to the CU within this range. However, a dramatic shift occurs as \(\alpha_c\) moves from 0.43 to around 0.5. With this change of $\alpha_c$, the CU's average BLER sharply rises, reaching nearly 1. This sudden increase suggests a critical threshold beyond which additional power allocation to the CU negatively impacts its BLER, likely due to over-allocation causing inefficiencies or other system limitations. Conversely, the CEU's average BLER starts very low when \(\alpha_c\) is close to 0, showcasing excellent performance with most power allocated to it. As \(\alpha_c\) increases to approximately 0.43, the BLER for the CEU gradually worsens but at a decelerating rate, indicating a resilience in performance despite the decreasing power. Once \(\alpha_c\) exceeds 0.5, the deterioration in the CEU’s average BLER accelerates, and then stabilizes near 1. These observations highlight the complex interplay between power allocation and user performance in a RIS-enhanced NOMA system. The graph provides crucial insights into the optimal power allocation ranges that balance the needs of both the CU and CEU, ensuring fair and efficient communication within the system. Such insights are vital for designing and implementing power control strategies in NOMA systems to maximize the overall network performance.

\begin{figure}[t] 
	\centering 
	\includegraphics[width=3.2in]{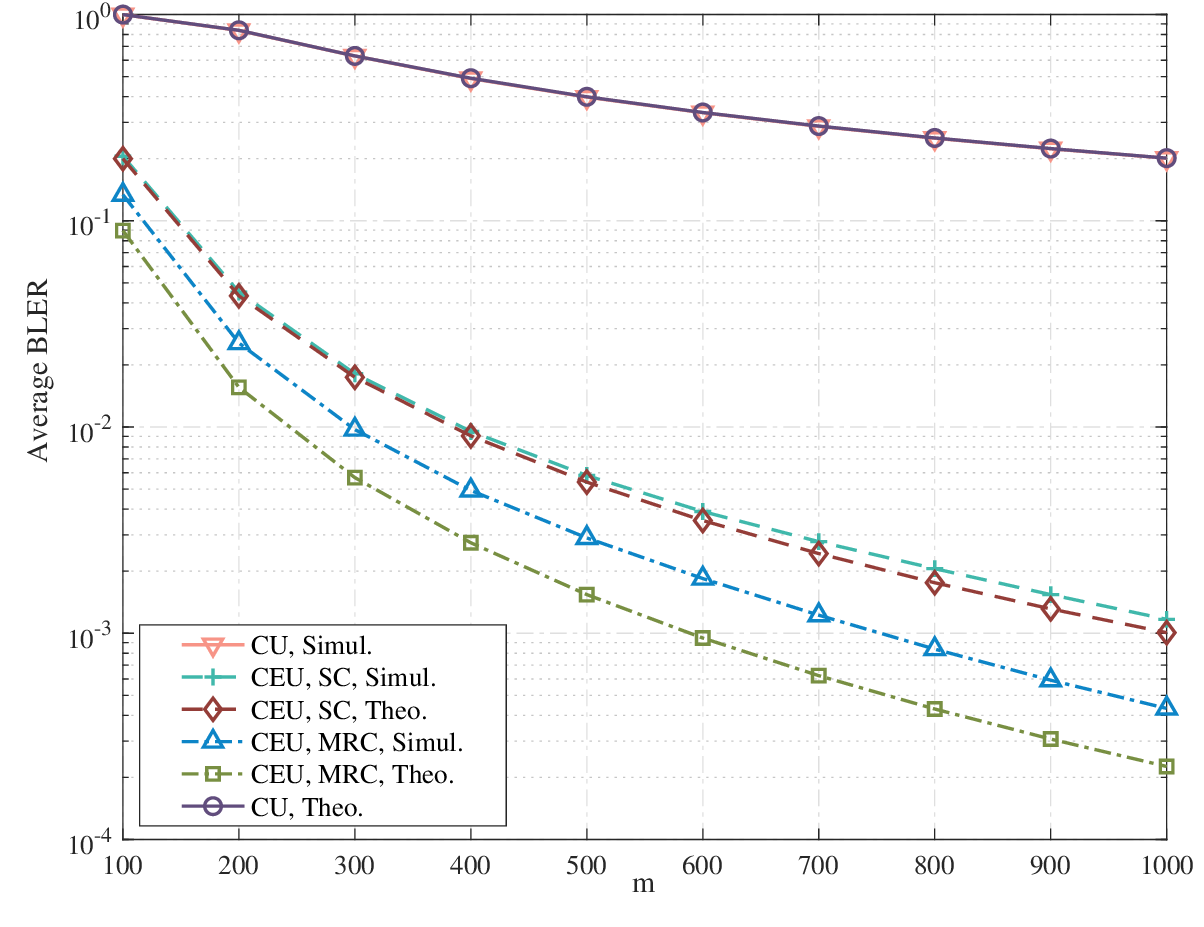}
	\caption{Average BLER versus $m$, where $R = 2$, $N_c = 300$ bits, $N_e = 100$ bits, $\rho_S=10$ dB, $\alpha_c = 0.1$, and $\alpha_e = 0.9$.}\label{Fig4}
\end{figure}

In Fig. \ref{Fig4}, we examine the average BLER of the CU and the CEU as a function of block length \(m\) in a setting where \(\rho_S=10\) dB, \(R=2\), \(\alpha_c = 0.1\), and \(\alpha_e = 0.9\), allocating minimal power to the CU and the majority to the CEU.  {Here, the channel use indicates the number of symbols used, and more symbols are used to represent the information bits sent, that is, the information is transmitted at a lower bit rate, which is suitable for scenarios with poor environments.} Fig. \ref{Fig4} shows that for both users, the average BLER decreases as \(m\) increases, indicating improved efficiency with larger data blocks. Notably, the SC scheme for both CU and CEU displays closely aligned simulated and theoretical results, confirming the accuracy of the theoretical models under the SC conditions.

\section{Conclusion}
In this paper, we have derived the average BLERs for a downlink cooperative NOMA URLLC communication system enhanced by RIS. Numerical results have been provided to demonstrate that the CU expeirenced consistent average BLER values across different configurations. For the CEU, the average BLER has remained nearly constant when SC has been employed, while a tight lower bound has been obtained under the MRC scheme. Furthermore, it has been evident that the proposed RIS-assisted cooperative NOMA system significantly outperforms its non-RIS counterpart, achieving a notably lower average BLER, which highlights the benefits of incorporating RIS for improving reliability. {Looking forward, significant research will focus on cooperative NOMA URLLC communication systems leveraging multiple-input multiple-output (MIMO) or fluid antenna systems (FAS).}

\appendices
\begin{figure*}[t]
	\centering 
	\begin{align}\label{q37}
		F_{\gamma_{cc}}\left(\omega\right)= \frac{\gamma\left(\kappa+1,\frac{\sqrt{\frac{\omega}{\alpha_c\rho_S}}}{\eta_c b_c}\right)}{\Gamma\left(\kappa+1\right)}-\frac{\sqrt{\frac{\omega}{\alpha_c\rho_S}}}{2\eta_cb_c^\kappa\Gamma\left(\kappa+1\right)}\exp\left(-\frac{\omega}{\alpha_c\rho_S\lambda_c}\right) 
		\cdot \sum_{u=1}^U\sqrt{1-\xi_u^2}\frac{\pi}{U}\zeta_{uc}^{\kappa}\left(\frac{\omega}{\alpha_c\rho_S}\right)\exp\left(\frac{\eta_c^2\zeta_{uc}^2\left(\frac{\omega}{\alpha_c\rho_S}\right)}{\lambda_c}-\frac{\zeta_{uc}\left(\frac{\omega}{\alpha_c\rho_S}\right)}{b_c}\right).
	\end{align}\vspace{3mm}
	\hrule 
	\begin{align}\label{q38}
		&\mathbb{E}\left[\epsilon_{cc}\right] \nonumber\\
		\approx&\frac{\gamma\left(\kappa+1,\frac{\sqrt{\frac{\beta_{N_c,m}}{\alpha_c\rho_S}}}{\eta_c b_c}\right)}{\Gamma\left(\kappa+1\right)}-\frac{\sqrt{\frac{\beta_{N_c,m}}{\alpha_c\rho_S}}}{2\eta_cb_c^\kappa\Gamma\left(\kappa+1\right)}\exp\left(-\frac{\beta_{N_c,m}}{\alpha_c\rho_S\lambda_c}\right) 
		\cdot \sum_{u=1}^U\sqrt{1-\xi_u^2}\frac{\pi}{U}\zeta_{uc}^{\kappa}\left(\frac{\beta_{N_c,m}}{\alpha_c\rho_S}\right)\exp\left(\frac{\eta_c^2\zeta_{uc}^2\left(\frac{\beta_{N_c,m}}{\alpha_c\rho_S}\right)}{\lambda_c}-\frac{\zeta_{uc}\left(\frac{\beta_{N_c,m}}{\alpha_c\rho_S}\right)}{b_c}\right).
	\end{align}  \vspace{3mm}	\hrule
	\begin{align}\label{q50}
		\mathbb{E}\left[\epsilon_{ce}\right]\approx\left\{
		\begin{array}{lcl}
			1, & & \beta_{N_e,m}\geq\frac{\alpha_e}{\alpha_c}\\
			\frac{\gamma\left(\kappa+1,\frac{\sqrt{\frac{\beta_{N_e,m}}{\alpha_e\rho_S-\alpha_c\rho_S \beta_{N_e,m}}}}{\eta_c b_c}\right)}{\Gamma\left(\kappa+1\right)}
			-\frac{\sqrt{\frac{\beta_{N_e,m}}{\alpha_e\rho_S-\alpha_c\rho_S \beta_{N_e,m}}}}{2\eta_cb_c^\kappa\Gamma\left(\kappa+1\right)}\exp\left(\frac{\beta_{N_e,m}}{\lambda_c\left(\alpha_e\rho_S-\alpha_c\rho_S \beta_{N_e,m}\right)}\right)\\
			\cdot\sum_{u=1}^U\sqrt{1-\xi_u^2}\frac{\pi}{U}\zeta_{uc}^{\kappa}\left(\frac{\beta_{N_e,m}}{\alpha_e\rho_S-\alpha_c\rho_S \beta_{N_e,m}}\right)\exp\left(\frac{\eta_c^2\zeta_{uc}^2\left(\frac{\beta_{N_e,m}}{\alpha_e\rho_S-\alpha_c\rho_S \beta_{N_e,m}}\right)}{\lambda_c}-\frac{\zeta_{uc}\left(\frac{\beta_{N_e,m}}{\alpha_e\rho_S-\alpha_c\rho_S \beta_{N_e,m}}\right)}{b_c}\right),& & \mbox{otherwise}
		\end{array} \right.
	\end{align} \vspace{2mm}
	\hrule 
	\begin{align}\label{q55}
		\mathbb{E}[\epsilon_{c}]\geq &\max \left\{\frac{\gamma\left(\kappa+1,\frac{\sqrt{\frac{\beta_{N_c,m}}{\alpha_c\rho_S}}}{\eta_c b_c}\right)}{\Gamma\left(\kappa+1\right)}-\frac{\sqrt{\frac{\beta_{N_c,m}}{\alpha_c\rho_S}}}{2\eta_cb_c^\kappa\Gamma\left(\kappa+1\right)}e^{\left(-\frac{\beta_{N_c,m}}{\alpha_c\rho_S\lambda_c}\right)}\right.  
		\cdot \sum_{u=1}^U\sqrt{1-\xi_u^2}\frac{\pi}{U}\zeta_{uc}^{\kappa}\left(\frac{\beta_{N_c,m}}{\alpha_c\rho_S}\right)e^{\left(\frac{\eta_c^2\zeta_{uc}^2\left(\frac{\beta_{N_c,m}}{\alpha_c\rho_S}\right)}{\lambda_c}-\frac{\zeta_{uc}\left(\frac{\beta_{N_c,m}}{\alpha_c\rho_S}\right)}{b_c}\right)},\nonumber\\
		&\Theta\left(\beta_{N_e,m}-\frac{\alpha_e}{\alpha_c}\right)+\Theta\left(\frac{\alpha_e}{\alpha_c}-\beta_{N_e,m}\right)\frac{\gamma\left(\kappa+1,\frac{\sqrt{\frac{\beta_{N_e,m}}{\alpha_e\rho_S-\alpha_c\rho_S \beta_{N_e,m}}}}{\eta_c b_c}\right)}{\Gamma\left(\kappa+1\right)}
		-\frac{\sqrt{\frac{\beta_{N_e,m}}{\alpha_e\rho_S-\alpha_c\rho_S \beta_{N_e,m}}}}{2\eta_cb_c^\kappa\Gamma\left(\kappa+1\right)}\exp\left(\frac{\beta_{N_e,m}}{\lambda_c\left(\alpha_e\rho_S-\alpha_c\rho_S \beta_{N_e,m}\right)}\right)\nonumber\\
		\cdot &\left.\sum_{u=1}^U\sqrt{1-\xi_u^2}\frac{\pi}{U}\zeta_{uc}^{\kappa}\left(\frac{\beta_{N_e,m}}{\alpha_e\rho_S-\alpha_c\rho_S \beta_{N_e,m}}\right)\exp\left(\frac{\eta_c^2\zeta_{uc}^2\left(\frac{\beta_{N_e,m}}{\alpha_e\rho_S-\alpha_c\rho_S \beta_{N_e,m}}\right)}{\lambda_c}-\frac{\zeta_{uc}\left(\frac{\beta_{N_e,m}}{\alpha_e\rho_S-\alpha_c\rho_S \beta_{N_e,m}}\right)}{b_c}\right)\right\}.
	\end{align}
	\hrule 
\end{figure*}
\section{Proof of Lemma 1}\label{A1}
Because of $h_c\sim \mathcal{CN}\left(0,\lambda_c\right)$, the PDF and CDF of $|h_c|^2$ are given by
\begin{align}\label{q97}
	f_{|h_c|^2}(x)&=\frac{1}{\lambda_c}\exp\left(-\frac{x}{\lambda_c}\right),\\
	\label{q98}F_{|h_c|^2}(x)&=1-\exp\left(-\frac{x}{\lambda_c}\right).
\end{align}

And the PDF and CDF of $q_c$ are given by\cite{LiS22}
\begin{align}\label{q99}
	f_{q_c}(y)&=\frac{y^{\kappa}}{b_c^\kappa\Gamma\left(\kappa+1\right)}\exp\left(-\frac{y}{b_c}\right),\\
	\label{q100}F_{q_c}(y)&=\frac{\gamma\left(\kappa+1,\frac{y}{b_c}\right)}{\Gamma\left(\kappa+1\right)}.
\end{align}

According to the definition of CDF, we have
\begin{align}\label{q101}
	F_T(t)=&\Pr\left(T\leq t\right)\nonumber\\
	=&\Pr\left(|h_c|^2\leq t-q_c^2\eta_c^2\right)
\end{align}
when $t\leq 0$,
\begin{align}\label{q102}
	\Pr\left(T\leq t\right)=0,
\end{align}
otherwise, $t\geq 0$ and $q_c \leq \frac{\sqrt{t}}{\eta_c}$, we have
\begin{align}\label{q103}
	F_T(t)=&\int_0^\frac{\sqrt{t}}{\eta_c}\int_0^{t-\eta_c^2 y^2}f_{|h_c|^2}(x)f_{q_c}(y)dxdy\nonumber\\
	=&\int_0^\frac{\sqrt{t}}{\eta_c}F_{|h_c|^2}\left(t-\eta_c^2y^2\right)f_{q_c}(y)dy.
\end{align}
Substituting \eqref{q98} and \eqref{q99} into \eqref{q103}, we have
\begin{align}\label{q104}
	F_T(t)=&\int_0^\frac{\sqrt{t}}{\eta_c}\left(1-\exp\left(-\frac{t-\eta_c^2y^2}{\lambda_c}\right)\right)f_{q_c}(y)dy\nonumber\\
	=&F_{q_c}\left(\frac{\sqrt{t}}{b_c\eta_c}\right)-\exp\left(-\frac{t}{\lambda_c}\right)\Upsilon,
\end{align}
where
\begin{align}\label{q105}
	\Upsilon=\int_0^\frac{\sqrt{t}}{\eta_c}\exp\left(\frac{\eta_c^2y^2}{\lambda_c}\right)f_{q_c}(y)dy,
\end{align}
Add \eqref{q99} into \eqref{q105}, we have
\begin{align}\label{aqx1}
	\Upsilon=\int_0^\frac{\sqrt{t}}{\eta_c} \frac{y^{\kappa}}{b_c^\kappa\Gamma\left(\kappa+1\right)}\exp\left(\frac{\eta_c^2y^2}{\lambda_c}-\frac{y}{b_c}\right)dy,
\end{align}
It is difficult to take the integral with a closed-form solution, we apply the Gaussian-Chebyshev quadrature of  \eqref{aqx1}.
\begin{align}\label{aqx2}
	\Upsilon=&\frac{\sqrt{t}}{2\eta_cb_c^\kappa\Gamma\left(\kappa+1\right)}\sum_{u=1}^U\sqrt{1-\xi_u^2}\frac{\pi}{U}\zeta_{uc}^{\kappa}(t)\nonumber\\
	\cdot&\exp\left(\frac{\eta_c^2\zeta_{uc}^2(t)}{\lambda_c}-\frac{\zeta_{uc}(t)}{b_c}\right).
\end{align}
Substituting \eqref{q100} and \eqref{aqx2} into \eqref{q103}, the CDF of $T$ can be obtained in \eqref{q32}.

\section{CDF of $\gamma_{cc}$}\label{A2}
The CDF of $\gamma_{cc}$ is given in \eqref{q37}. 

\section{Proof of Lemma 2}\label{A3}
The SNR $\gamma_{cc}$ can be written as
\begin{align}\label{q108}
	\gamma_{cc}=\alpha_c\rho_S T,
\end{align}
the CDF of $\gamma_{cc}$ is given by
\begin{align}\label{q109}
	F_{\gamma_{cc}}\left(\omega\right)=&\Pr\left(\gamma_{cc}\leq \omega\right) 
\end{align}
add \eqref{q109} into \eqref{q108}, we have
\begin{align}\label{q110}
	F_{\gamma_{cc}}\left(\omega\right)=&\Pr\left(T\leq \frac{\omega}{\alpha_c\rho_S}\right)\nonumber\\
	=&F_T\left(\frac{\omega}{\alpha_c\rho_S}\right),
\end{align}
and add $\frac{\omega}{\alpha_c\rho_S}$ into \eqref{q32}, we have Lemma 2.

\begin{figure*}[t] 
	\centering
	\vspace{-10mm}
	\begin{align}\label{q76}
		&\mathbb{E}\left[\Psi(\gamma_{e},N_e,m)\right]\nonumber\\
		\approx& \left(\Theta\left(\beta_{N_e,m}-\frac{\alpha_e}{\alpha_c}\right)+\Theta\left(\frac{\alpha_e}{\alpha_c}-\beta_{N_e,m}\right)\frac{\gamma\left(\kappa+1,\frac{\sqrt{\frac{\beta_{N_e,m}}{\alpha_e\rho_S-\alpha_c\rho_S\beta_{N_e,m}}}}{\eta_e b_e}\right)}{\Gamma\left(a_e+1\right)}-\frac{\sqrt{\frac{\beta_{N_e,m}}{\alpha_e\rho_S-\alpha_c\rho_S\beta_{N_e,m}}}}{2\eta_eb_e^\kappa\Gamma\left(\kappa+1\right)}\exp\left(-\frac{\beta_{N_e,m}}{\lambda_e\left(\alpha_e\rho_S-\alpha_c\rho_S\beta_{N_e,m}\right)}\right)\right.\nonumber\\
		\cdot&\left.\sum_{u=1}^U\sqrt{1-\xi_u^2}\frac{\pi}{U}\zeta_{ue}^{\kappa}\left(\frac{\beta_{N_e,m}}{\alpha_e\rho_S-\alpha_c\rho_S\beta_{N_e,m}}\right)\exp\left(\frac{\eta_e^2\zeta_{ue}^2\frac{\beta_{N_e,m}}{\alpha_e\rho_S-\alpha_c\rho_S\beta_{N_e,m}}}{\lambda_e}-\frac{\zeta_{ue}\frac{\beta_{N_e,m}}{\alpha_e\rho_S-\alpha_c\rho_S\beta_{N_e,m}}}{b_e}\right)\right)\nonumber\\
		\cdot&\left(\frac{\gamma\left(\kappa+1,\frac{\sqrt{\frac{\beta_{N_e,m}}{\rho_C}}}{\eta_e b_e}\right)}{\Gamma\left(\kappa+1\right)}\frac{\sqrt{\frac{\beta_{N_e,m}}{\rho_C}}}{2\eta_eb_{ce}^\kappa\Gamma\left(\kappa+1\right)}\exp\left(-\frac{\beta_{N_e,m}}{\rho_C\lambda_e}\right)\right.\nonumber\\
		\cdot&\left.\sum_{u=1}^U\sqrt{1-\xi_u^2}\frac{\pi}{U}\zeta_{ue}^{\kappa}\left(\frac{\beta_{N_e,m}}{\rho_C}\right)\exp\left(\frac{\eta_e^2\zeta_{ue}^2\left(\frac{\beta_{N_e,m}}{\rho_C}\right)}{\lambda_e}-\frac{\zeta_{ue}\left(\frac{\beta_{N_e,m}}{\rho_C}\right)}{b_{ce}}\right)\right).
	\end{align}	\hrule
	\begin{align}\label{q81}
		&\mathbb{E}\left[\Psi(2\gamma_{e1},N_e,m)\right]\nonumber\\
		\approx& \left(\Theta\left(\beta_{N_e,m}-\frac{\alpha_e}{\alpha_c}\right)+\Theta\left(\frac{\alpha_e}{\alpha_c}-\beta_{N_e,m}\right)\frac{\gamma\left(\kappa+1,\frac{\sqrt{\frac{\beta_{N_e,m}}{2\left(\alpha_e\rho_S-\alpha_c\rho_S\beta_{N_e,m}\right)}}}{\eta_e b_e}\right)}{\Gamma\left(a_e+1\right)}\right.\nonumber\\
		-&\frac{\sqrt{\frac{\beta_{N_e,m}}{2\left(\alpha_e\rho_S-\alpha_c\rho_S\beta_{N_e,m}\right)}}}{2\eta_eb_e^\kappa\Gamma\left(\kappa+1\right)}\exp\left(-\frac{\beta_{N_e,m}}{2\lambda_e\left(\alpha_e\rho_S-\alpha_c\rho_S\beta_{N_e,m}\right)}\right)\nonumber\\
		\cdot&\left.\sum_{u=1}^U\sqrt{1-\xi_u^2}\frac{\pi}{U}\zeta_{ue}^{\kappa}\left(\frac{\beta_{N_e,m}}{2\left(\alpha_e\rho_S-\alpha_c\rho_S\beta_{N_e,m}\right)}\right)\exp\left(\frac{\eta_e^2\zeta_{ue}^2\frac{\beta_{N_e,m}}{2\left(\alpha_e\rho_S-\alpha_c\rho_S\beta_{N_e,m}\right)}}{\lambda_e}-\frac{\zeta_{ue}\frac{\beta_{N_e,m}}{2\left(\alpha_e\rho_S-\alpha_c\rho_S\beta_{N_e,m}\right)}}{b_e}\right)\right).
	\end{align}	\hrule
\end{figure*}
\section{Expressions of expectation}\label{A4}
 Expression of $\mathbb{E}\left[\epsilon_{cc}\right]$ is given by \eqref{q38}.

 Expression of $\mathbb{E}\left[\epsilon_{ce}\right]$ is given by \eqref{q50}.
 
Expression of $\mathbb{E}[\epsilon_{c}]$ is given by \eqref{q55}.

Expression of $\mathbb{E}\left[\Psi(\gamma_{e},N_e,m)\right]$ is given by \eqref{q76}.

Expression of  $\mathbb{E}[\Psi(2\gamma_{e1},N_e,m)]$ is given  by \eqref{q81}. 

\section{Proof of Lemma 3}\label{A5}
The SINR $\gamma_{ce}$ can be written as
\begin{align}\label{q111}
	\gamma_{ce}=\frac{\alpha_c\rho_S T}{\alpha_e\rho_S T+1},
\end{align}
the CDF of $\gamma_{ce}$ is given by
\begin{align}\label{q112}
	F_{\gamma_{ce}}\left(\omega\right)=&\Pr\left(\gamma_{ce}\leq \omega\right)\nonumber\\
	=&\Pr\left(T\leq \frac{\omega}{\alpha_e\rho_S-\alpha_c\rho_S \omega}\right),
\end{align}
when $\omega\geq\frac{\alpha_e}{\alpha_c}$,
\begin{align}\label{q113}
	F_{\gamma_{ce}}(\omega)=1,
\end{align}
otherwise, 
\begin{align}\label{q114}
	F_{\gamma_{ce}}(\omega)=&F_T\left(\frac{\omega}{\alpha_e\rho_S-\alpha_c\rho_S \omega}\right),
\end{align}
and add $\frac{\omega}{\alpha_e\rho_S-\alpha_c\rho_S \omega}$ into \eqref{q114}, we have Lemma 3.

\end{document}